# Universal Ratio of Intrinsic Resistivities of Spin Helix in B20 (Fe-Co)Si Magnets

S. X. Huang[1*,#], Jian Kang[2], Fei Chen[1,3], Jiadong Zang[1*], G. J. Shu[4], F. C. Chou[4], S. V. Grigoriev[5], V. A. Dyadkin[6], and C. L. Chien[1*]

[1]*Department of Physics and Astronomy, Johns Hopkins University, Baltimore, MD 21218, USA*

[2]*School of Physics and Astronomy, University of Minnesota, Minneapolis, MN 55455, USA*

[3]*Department of physics, Nanjing University, Nanjing 210093, China*

[4]*Center for Condensed Matter Sciences, National Taiwan University, Taipei 10617, Taiwan*

[5]*Petersburg Nuclear Physics Institute, 188300 Gatchina, Saint Petersburg, Russia*

[6]*Swiss-Norwegian Beamlines at the ESRF, 38000 Grenoble, France*

Abstract:

The B20 magnets with the Dzyaloshinskii-Moriya (D-M) interaction exhibit spin helix and Skyrmion spin textures unattainable in traditional Heisenberg ferromagnets. We have determined the intrinsic resistivity of the spin helix, which is a macroscopic Bloch domain wall, in B20 (Fe-Co)Si magnets. We found a universal resistance ratio of $\gamma \approx 1.35$ with current parallel and perpendicular to the helix, independent of composition and temperature. This $\gamma$ value is much smaller than 3, the well-known minimum value for domain wall resistivity in traditional ferromagnets, due to the significant spin-orbit coupling in the B20 magnets.

One of the cornerstones of spintronics is spin-dependent scattering of conduction electrons in ferromagnetic (FM) heterostructures as first revealed in giant magnetoresistance (GMR) [1, 2]. There are also resistive effects in a single FM metal. The most well-known is the anisotropic MR (AMR), which depends on the angle between the magnetization *M* (as aligned by a magnetic field) and the electrical current *I*. In a FM metal with magnetic domains, there is also domain wall resistance (DWR), [3-12] which depends on the intricate spin structure (e.g., Néel type and Bloch type) within the tiny domain wall (DW).

The aforementioned effects of GMR, AMR, and DWR occur in traditional FM metals (e.g., Fe and Co) with a FM ground state due to the Heisenberg exchange interaction with energy $A(\nabla M)^2$ that favors parallel moment alignment, where *A* is the exchange stiffness constant and *M* is the magnetization. Recently, a new type of B20 chiral magnets (MnSi, FeGe, $Fe_{1-v}Co_v Si$ etc.) *without* inversion symmetry has attracted a great deal of attention [13-15]. The broken inversion symmetry and the spin-orbit coupling give rise to the nontrivial Dzyaloshinskii-Moriya (D-M) interaction[16,17], with energy $D M \times (\nabla \times M)$, that favors perpendicular moment alignment, where *D* is the Dzyaloshinskii constant. The competition between the D-M and Heisenberg interactions leads to a unique spin helix ground state [18-20]. Under a magnetic field, there are also the conical state and especially the Skyrmion state as revealed by neutron diffraction and Lorentz microscopy [14, 15], with a host of spectacular dynamic properties [21, 22]. These new spin structures, hitherto unavailable in traditional Heisenberg FMs, provide new media for interaction with the conduction electrons, in particular the intrinsic resistance due to the new spin structures.

In the spin helix in the B20 magnets, the spins are aligned in successive planes, but the spin orientation advances by a fixed angle along the spin helix propagation direction $k_H$, with a spin helix wavelength of $\lambda_H = 4\pi A/D$. In contrast to the small DWs in traditional FM materials, the spin helix is a *macroscopic* Bloch DW extending throughout the sample. We report in this work the determination of the intrinsic resistance of spin helix in single crystals of $Fe_{1-v}Co_vSi$ ($v$=0.15, 0.3 and 0.5) with current parallel and perpendicular to the helix and the resistance ratio γ. We have observed a *universal* resistance ratio of γ ≈ 1.35, independent of composition and temperature, and much smaller than all those observed in the DWR in traditional FM. We show this is due to the spin-orbit coupling contribution inherent to the D-M interaction.

The hallmarks of traditional ferromagnets with a FM ground state are anisotropy, hysteresis, and remnant magnetization. The B20 magnets, with a spin helical ground state, are diametrically different. The hysteresis M-H curve of $Fe_{0.7}Co_{0.3}Si$ at 5 K shows no loop width nor remnant magnetization as shown in Fig. 1a. The loops with the field along $z$[001], $y$[110], and other directions are essentially the same. Negligible magnetic anisotropy, non-hysteric and zero remnant magnetization are the characteristics of spin helices in B20 magnets.

The remnant state with zero magnetization is due to the spin helical state with $k_H$ along the original field direction. Starting from the spin helix state, with increasing field, a conical spin structure is formed until all the spins are aligned at the saturation field ($H_S$). These spin structures are schematically shown in Fig. 1a. At $H$ = 2 kOe, which is above $H_S$, the resistivity with field in the $xy$ and the $yz$ planes shows a $(cosine)^2$ angular dependence as shown in Fig. 1b. Because of the negligible anisotropy, the helical wave vector $k_H$ can be installed by the direction of an applied magnetic field. This feature in $Fe_{1-v}Co_vSi$ has been demonstrated by small angle neutron scattering (SANS) measurements [23]. $k_H$ remains in the original field direction after the field has been removed [23]. Thus, the application of a magnetic field ($H > H_S$) defines a new direction for the formation of the spin helix. This unique feature is revealed in the angular dependence of resistivity of the remnant state in *zero field* by first applying $H > H_S$ at each angle and measuring $R$ after the field has been turned off. As shown in Fig. 1c, although the magnetization is zero at all angles, the resistivity at zero field also shows a (cosine)$^2$ angular dependence, confirming the spin helix at different angles.

Before further discussion we should mention that it has been known that the B20 magnets including $Fe_{1-y}Co_ySi$, exhibit a positive linear MR, which remains unabated to very high field in excess of 100 kOe[24]. This intriguing and poorly understood effect, independent on the field direction[24, 25], may have a complex origin including quantum interference effects[24] and the Zeeman shift of the exchange-split bands[26]. In the following, we subtract this linear MR background, which is quite small in the low field ranges of less than 3 kOe.

In Fig. 2a, we show the field dependence of resistivity at 3 K of $Fe_{0.85}Co_{0.15}Si$ single crystal S1 with sample edges parallel to the cubic axes. Below $H_S$, the field dependence of MR closely correlates with the M-H curves with the same $H_S$ indicating that the MR below $H_S$ originates from the helical and conical spin textures. Above $H_S$ with all the spins aligned, the resistivities $\rho_y$ and $\rho_z$ with field along $y$ and $z$ axes respectively are *not* the same, a clear departure from the usual AMR with cubic symmetry. This is the consequence of the broken $C_4$ symmetry unique to the B20 magnets [25]. As a result, the angular scan of the MR with field in *xy*, *yz*, or *zx* planes show 2-fold and not 4-fold symmetry (Fig. 2c). One notes that the angular dependence at zero field (Fig. 1c) also have 2-fold symmetry for the same reason. For $Fe_{0.85}Co_{0.15}Si$-S1, the resistance maxima and minima are located at multiples of 90°. We have also measured another $Fe_{0.85}Co_{0.15}Si$ (S2) crystal with the sample edges deliberately off the crystal axes. The field dependence of S2 (Fig. 2b) is much different from that of S1 (Fig. 2a). The offset is clearly revealed in the angular scan shown in Fig. 2d, but the 2-fold symmetry remains.

The field dependences of the resistances of $Fe_{0.85}Co_{0.15}Si$-S1, measured with the magnetic field along the *x, y, z,* axes, as shown in Fig. 2a, exhibit 6 characteristic values. The values of $\rho_x$, $\rho_y$ and $\rho_z$ (denoted as colored solid circles in Fig. 2a) at large fields ($H > H_S$) are those with aligned spins along the *x, y,* and *z* axes respectively. These are also the maximal and minimal values in the angular scans shown in Fig. 2c. The values of $\rho_{x0}$, $\rho_{y0}$, and $\rho_{z0}$ (denoted as colored open circles in Fig. 2a) are those at $H = 0$, *after* first applying a large magnetic field, thus with a spin helix formed, along the *x, y,* and *z* axes respectively.

We now discuss the determination of the intrinsic spin helix resistance, which characterizes the collective contribution of interaction between conducting electrons and the spin helix. The value of $\rho_{x0}$ contains the contribution of the spin helix in the *x*[100] direction and the AMR contribution with spins in the *yz* plane. However, because in $\rho_y$ and $\rho_z$ are not the same due to the broken $C_4$ symmetry, the AMR contribution needs to be determined and subtracted from $\rho_{x0}$. With *H* in the *yz* plane, as shown by the blue curve in Fig.2c, the resistivity $\rho_{yz}(\theta)$ in the yz plane has an angular dependence of $\rho_{yz}(\theta) = \rho_y + (\rho_z - \rho_y)\cos^2\theta$, which gives rise to an AMR contribution by spin helix of

$$\rho_{yz}^{AMR} = \int_0^{2\pi} \rho_{yz}(\theta)d\theta/2\pi = (\rho_z + \rho_y)/2. \tag{1}$$

Thus the intrinsic resistivity of spin helix in the *x*-direction is

$$\rho_x^{Helix} = \rho_{x0} - \rho_{yz}^{AMR} = \rho_{x0} - (\rho_z + \rho_y)/2. \tag{2}$$

In a similar manner, the intrinsic resistivities in the *y* and *z* directions are $\rho_y^{Helix} = \rho_{y0} - (\rho_z + \rho_x)/2$ and $\rho_z^{Helix} = \rho_{z0} - (\rho_x + \rho_y)/2$. Since in all three cases, the current is applied along the *x* direction, $\rho_x^{Helix}$ is the intrinsic helix resistance with the current parallel to the helix, whereas both $\rho_y^{Helix}$ and $\rho_z^{Helix}$ are those with the current perpendicular to the helix. With the six experimental values of $\rho_{x0}$, $\rho_{y0}$, and $\rho_{z0}$ of 354.56 μΩ-cm, 355.95 μΩ-cm, and 354.81 μΩ-cm respectively, and $\rho_x$, $\rho_y$, and $\rho_z$ of 355.06 μΩ-cm, 350.68 μΩ-cm, and 353.04 μΩ-cm respectively, we obtain the numerical values of $\rho_x^{Helix}$ = 2.70±0.1 μΩ·cm, $\rho_y^{Helix}$ = 1.90±0.1 μΩ·cm, and $\rho_z^{Helix}$ = 1.94±0.1 μΩ·cm for Fe$_{0.85}$Co$_{0.15}$Si (S1) at 3 K. All the spin helix resistances are positive and with $\rho_x^{Helix} > \rho_y^{Helix}$ and $\rho_y^{Helix} \approx \rho_z^{Helix}$. The latter result is expected because in both cases the current is perpendicular to the spin helix direction.

We can generalize this method to include samples whose edges are off the crystalline axes as in the sample Fe$_{0.85}$Co$_{0.15}$Si (S2). From the field dependence curves shown in Fig. 2b, the values of $\rho_{x0}$, $\rho_{y0}$, and $\rho_{z0}$ at $H = 0$ are useful but not the values of $\rho_x$, $\rho_y$, and $\rho_z$ at $H > H_S$ because they are not the extremal values. Instead, as shown in Fig. 2d, each of the three angular scans shows a maximal and a minimal value that are needed for calculating the AMR contribution. Considering a spin helix with $k_H$ ($k_H = x$, $y$ or $z$) direction perpendicular to the $ij$ plane (that is, the spins are distributed uniformly in the $ij$ plane), the resistivity due to AMR contribution is then

$$\rho_{ij}^{AMR} = \int_0^{2\pi} \rho_{ij}(\theta_{ij}) d\theta_{ij} / 2\pi = (\rho_{ij}^{\min} + \rho_{ij}^{\max})/2. \qquad (3)$$

The intrinsic resistivity $\rho_{k(helix)}$ of the spin helix can be determined by $\rho_k^{Helix} = \rho_{k0} - \rho_{ij}^{AMR}$, where $\rho_{k0}$ is the experimentally measured resistivity at zero field. In this manner, for Fe$_{0.85}$Co$_{0.15}$Si (S2) at 3 K, we have determined similar values of $\rho_x^{Helix}$ = 2.65 μΩ·cm, $\rho_y^{Helix}$ = 2.01 μΩ·cm, $\rho_z^{Helix}$ = 2.11 μΩ·cm. Despite the different crystalline orientations for Fe$_{0.85}$Co$_{0.15}$Si

S1 and S2 (thus different field and angular dependence of MR), the intrinsic resistivities of spin helix remain the same.

As temperature increases, the values of the intrinsic spin helix resistivities decrease and vanish at $T_C$ as expected but the feature of $\rho_x^{Helix} > \rho_y^{Helix} \approx \rho_z^{Helix}$ remains unchanged. Of particularly interest is the resistivity ratio $\gamma = 2\rho_x^{Helix}/(\rho_y^{Helix} + \rho_z^{Helix})$, which is approximately $\rho_x^{Helix}/\rho_y^{Helix}$, a measure of the difference in scattering between current parallel and perpendicular to the spin helix direction. Remarkably, the ratio of $\gamma \approx 1.35$ has been consistently observed and found to be independent of temperature, even when the intrinsic resitivities of spin helix change by about a factor of three from 3 K to 21 K as shown in Fig. 3b. Furthermore, we have also measured the intrinsic spin helix for other compositions, including $Fe_{0.7}Co_{0.3}Si$ ($\rho$ = 288 µΩ·cm at 3 K) with $\rho_x^{Helix}$ =1.16 µΩ·cm and $Fe_{0.5}Co_{0.5}Si$ ($\rho$ = 224 µΩ·cm at 15 K) with $\rho_x^{Helix}$ = 0.11µΩ·cm; both are much smaller than $\rho_x^{Helix}$ = 2.7 µΩ·cm in $Fe_{0.85}Co_{0.15}Si$. Thus, even when the intrinsic resistivity of spin helix $\rho_x^{Helix}$ varies by 25 times, the resistivity ratio $\gamma$ remains the same at 1.35. These results indicate that $\gamma$ = 1.35 is *universal* for spin helix in $Fe_{1-v}Co_vSi$.

The measured intrinsic resistivity is positive, showing increased scattering in the presence of a spin helix. A spin helix has the appearance of a macroscopic Bloch domain wall, but the spin helix resistance in the B20 magnets is very different from the DWR in traditional FM metals. The Levy-Zhang's theory on DWR, including only spin dependent scattering, predicts a DWR resistivity ratio of $\gamma$ larger than a minimum value of about 3 [5], which has been observed in traditional FM materials [9]. In contrast, the resistivity ratio $\gamma$ is only 1.35 in the spin helix resistance in $Fe_{1-v}Co_vSi$ even though the helical wavelength in B20 magnets and the DW size in traditional FM materials are comparable. The key difference is the spin-orbit interaction, which plays a central role in the D-M interaction in B20 magnets, as well as its transport properties.

For the electronic transport, the Hamiltonian given by $H = \frac{\hbar^2 \mathbf{k}^2}{2m} - J_H \mathbf{M}(\mathbf{r}) \cdot \boldsymbol{\sigma}$, where $m$ is the effective mass and $J_H$ is the Hund's rule coupling, has been well established in the theories of GMR. A well-known microscopic D-M interaction has the form of $\mathbf{D}_{ij} \cdot (\mathbf{S}_i \times \mathbf{S}_j)$ between two neighboring spins $\mathbf{S}_i$ and $\mathbf{S}_j$ at locations $\mathbf{r}_i$ and $\mathbf{r}_j$ respectively, with $\mathbf{D}_{ij} \parallel (\mathbf{r}_i - \mathbf{r}_j)$. The D-M interaction of $\mathbf{D}_{ij} \cdot (\mathbf{S}_i \times \mathbf{S}_j)$ compels the two spins $\mathbf{S}_i$ and $\mathbf{S}_j$ to be perpendicular to each other and in a plane perpendicular to $\mathbf{D}_{ij}$. When an electron of spin direction $\boldsymbol{\sigma}$ hops from $\mathbf{S}_i$ to $\mathbf{S}_j$, its spin must rotate, hence effectively precesses about its momentum direction $\mathbf{k}$, as if under a magnetic field applied along the $\mathbf{k}$ direction. Thus, the spin-orbit coupling term in the electron transport compatible with the D-M interaction has the form of $\mathbf{k} \cdot \boldsymbol{\sigma}$. Consequently the full effective Hamiltonian of the conduction electron is given by $H = \frac{\hbar^2 \mathbf{k}^2}{2m} - \alpha \mathbf{k} \cdot \boldsymbol{\sigma} - J_H \mathbf{M}(\mathbf{r}) \cdot \boldsymbol{\sigma}$, where $\alpha$ is the parameter denoting the strength of the spin-orbit contribution. In the limit of $\alpha = 0$, one has the last term in Eq. (8) for spin-dependent scattering. The spin-orbit term of $\alpha \mathbf{k} \cdot \boldsymbol{\sigma}$ has interesting consequences. It promotes spin orientation $\boldsymbol{\sigma}$ to be *parallel* to the electron motion $\mathbf{k}$. Thus the spin-orbit term, when dominates, gives *isotropic* transport of $\gamma = 1$. This is the fundamental reason why $\gamma \approx 1.35$, substantially lower than 3, has been observed in the spin helix.

We have calculated the helix resistivity by the Boltzmann approach, the details of which will be published elsewhere. Here we discuss only the results of the resistivity ratio. The calculated relation between the resistivity ratio $\gamma$ and the strength of spin-orbit coupling, scaled as $\alpha k_F / J_H$, is shown in Fig. 4, where $k_F$ is the Fermi momentum. At $\alpha = 0$ the value of ratio $\gamma =$ 3.8 is in agreement with the Levy-Zhang theory. The value of $\gamma$ decreases sharply by the spin-orbit coupling. The value of $\gamma$ approaches 1 at large values of $|\alpha|$ as anticipated. It is interesting to note that $\gamma$ is not symmetric in $\alpha$ due to the following reason. When an electron traverses along a spin helix, its spin follows the local spiraling moments, resembling spin precession under an effective field $h_{eff}$ along the helix propagation direction. The time $T$ electron traverses through a helix period $T = m\lambda_H / \hbar k$ should equal to the precession period $2\pi\hbar / h_{eff}$. Consequently $h_{eff} = 2\pi\hbar^2 k / m\lambda_H$. The anisotropy of this effective field is closely related to that of the helix resistance. When $\alpha$ is positive, helicity of the conduction electron matches the spin helix, $h_{eff}$ and the effective field $\alpha k$ raised by spin-orbit coupling are additive, so that the degree of anisotropy decreases monotonously and slowly with $\alpha$. However once $\alpha$ is negative, $\alpha k$ cancels portion of $h_{eff}$, leading to a rapid drop of resistivity ratio $\gamma$. At $\alpha \approx -2\pi\hbar^2 / m\lambda_H$, $\gamma$ reaches its minimum value as shown in Fig. 4.

This physical picture is confirmed quantitatively. The large resistivity ratio is brought down by the spin-orbit coupling $\alpha$. Numerically, as shown in Fig. 4, the experimental value of $\gamma$ = 1.35 can be accommodated by $\alpha$ of *-0.68$J_H$* or *1.2$J_H$*, where $J_H$ is usually about *0.5* eV. This simple analysis does not determine the sign of spin-orbit coupling, which awaits density functional studies.

In summary, we have determined the intrinsic resistivities of spin helix for current parallel and perpendicular to spin helix. The intrinsic resistivity is as high as 2.7 µΩ·cm, much higher than the DWR in ordinary ferromagnets. We found a universal resistivity ratio $\gamma \approx 1.35$ in $Fe_{1-v}Co_vSi$, independent of temperature and compositions. Our calculations show this smaller value than those of DWR in traditional Heisenberg ferromagnets is a direct consequence of the spin-orbit interaction in these B20 magnets.

Work supported by NSF Grant DMR-1262253. SXH was partially supported by STARnet. J. Z. is supported by the Theoretical Interdisciplinary Physics and Astrophysics Center and by the US DOE DEFG02-08ER46544. We thank Professors Oleg Tchernyshyov and Rafael M. Fernandes for helpful discussions.

Figure captions:

Fig. 1 (color online). (a) Magnetic hysteresis loops of $Fe_{0.7}Co_{0.3}Si$ for field along [110] (red) and [001] (blue) directions. Inset: Spin structures of aligned state ($H>H_S$), conical state ($H<H_S$) and helical state ($H=0$); Resistivity of $Fe_{0.85}Co_{0.15}Si$-S1 in (b) the aligned state ($H = 2$ kOe $>H_S$), and (c) the remanence state as a function of angle with field in the $xy$ and $yz$ planes.

Fig. 2 (color online). Resistivity of (a) $Fe_{0.85}Co_{0.15}Si$-S1 and (b) $Fe_{0.85}Co_{0.15}Si$-S2 as a function of field (sweep from $H>H_S$) at 3K; Resistivity as a function of field angle ($T=3K$, $H=2$ kOe $> H_S$) for (c) $Fe_{0.85}Co_{0.15}Si$-S1 and (d) $Fe_{0.85}Co_{0.15}Si$-S2 with field lying in the $xy$ plane (black), $xz$ plane (red) and $yz$ plane (blue), respectively. Solid lines are $\cos^2\theta$ fit to data.

Fig. 3 (color online). (a) Spin helix with $k_H$ along $x$, $y$, and $z$ directions. (b) Intrinsic resistivities for $k_H$ along $x$ (black square symbols), $y$ (red circle symbols), and $z$ (blue triangle symbols) directions, and resistivity ratio (green circle symbols, see text for definition) in $Fe_{0.85}Co_{0.15}Si$ as a function of temperature, respectively. Solid symbols are for $Fe_{0.85}Co_{0.15}Si$-S1. Open symbols are for $Fe_{0.85}Co_{0.15}Si$-S2. (c) Intrinsic resistivities and resistivity ratio as a function of $v$. The temperature $T$ is 3K, 3K, 15K for $v=0.15$ ($T_C\approx23K$), $v=0.3$ ($T_C\approx48K$), and $v=0.5$ ($T_C\approx40K$), respectively. Solid lines ($\gamma =1.35$) are guides to eyes.

Fig. 4 (color online). Resistivity ratio as a function of $\alpha k_F/J_H$.

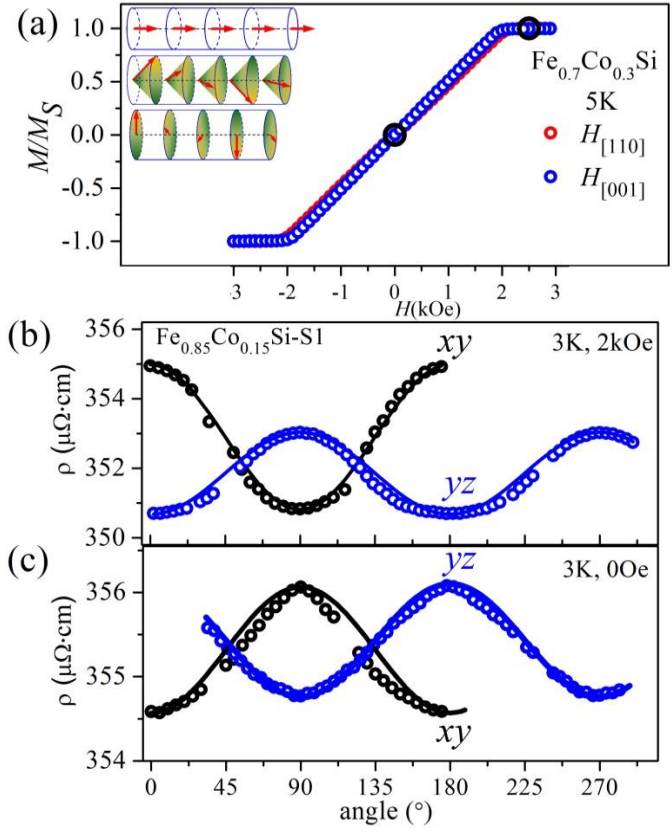

Fig. 1 (color online). (a) Magnetic hysteresis loops of $Fe_{0.7}Co_{0.3}Si$ for field along [110] (red) and [001] (blue) directions. Inset: Spin structures of aligned state ($H>H_S$), conical state ($H<H_S$) and helical state ($H=0$); Resistivity of $Fe_{0.85}Co_{0.15}Si$-S1 in (b) the aligned state ($H = 2$ kOe $>H_S$), and (c) the remanence state as a function of angle with field in the *xy* and *yz* planes.

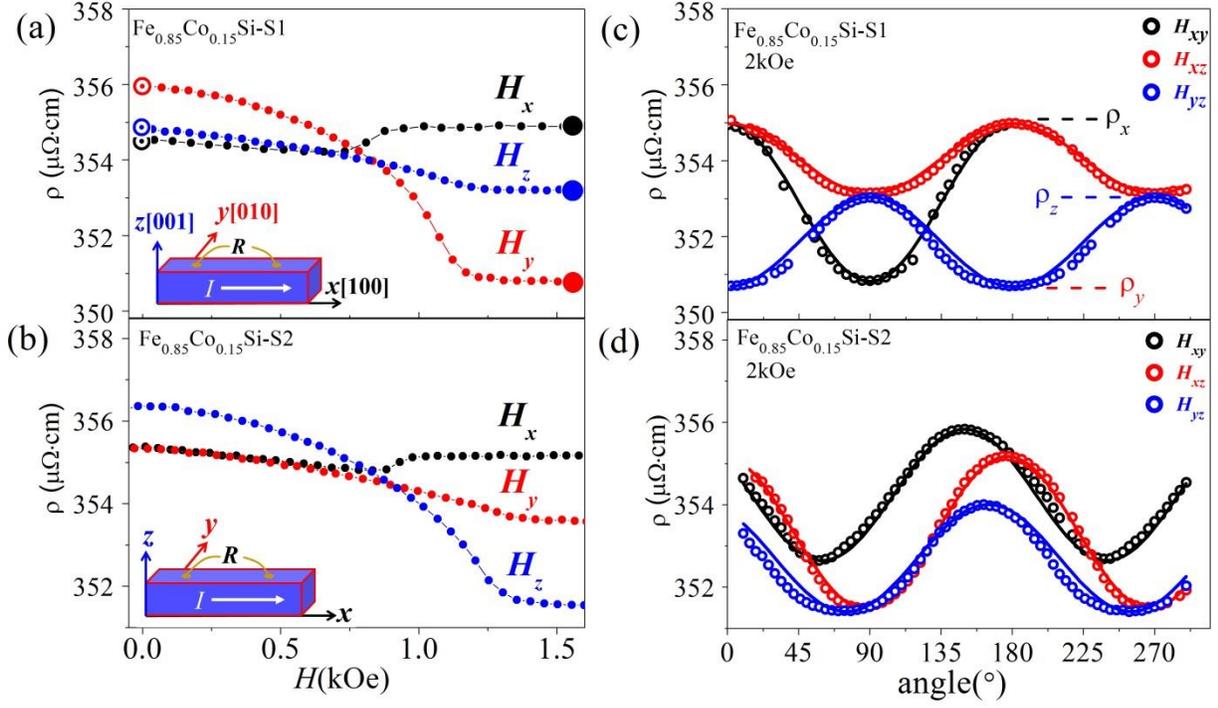

Fig. 2 (color online). Resistivity of (a) $Fe_{0.85}Co_{0.15}Si$-S1 and (b) $Fe_{0.85}Co_{0.15}Si$-S2 as a function of field (sweep from $H>H_S$) at 3K; Resistivity as a function of field angle ($T=3K$, $H=2$ kOe $> H_S$) for (c) $Fe_{0.85}Co_{0.15}Si$-S1 and (d) $Fe_{0.85}Co_{0.15}Si$-S2 with field lying in the $xy$ plane (black), $xz$ plane (red) and $yz$ plane (blue), respectively. Solid lines are $\cos^2\theta$ fit to data.

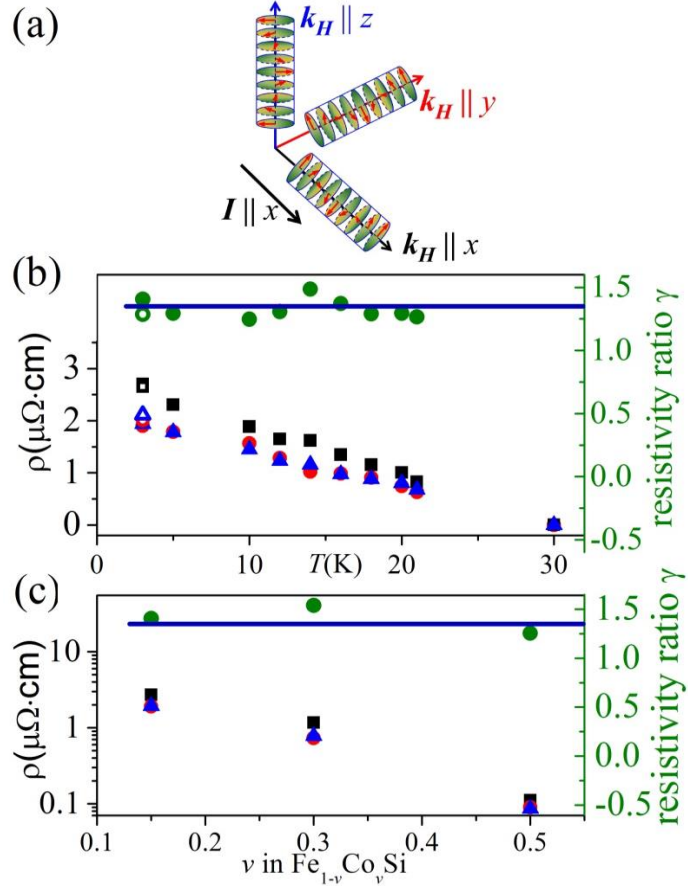

Fig. 3 (color online). (a) Spin helix with $k_H$ along $x$, $y$, and $z$ directions. (b) Intrinsic resistivities for $k_H$ along $x$ (black square symbols), $y$ (red circle symbols), and $z$ (blue triangle symbols) directions, and resistivity ratio (green circle symbols, see text for definition) in $Fe_{0.85}Co_{0.15}Si$ as a function of temperature, respectively. Solid symbols are for $Fe_{0.85}Co_{0.15}Si$-S1. Open symbols are for $Fe_{0.85}Co_{0.15}Si$-S2. (c) Intrinsic resistivities and resistivity ratio as a function of $v$. The temperature $T$ is 3K, 3K, 15K for $v=0.15$ ($T_C\approx23K$), $v=0.3$ ($T_C\approx48K$), and $v=0.5$ ($T_C\approx40K$), respectively. Solid lines ($\gamma=1.35$) are guides to eyes.

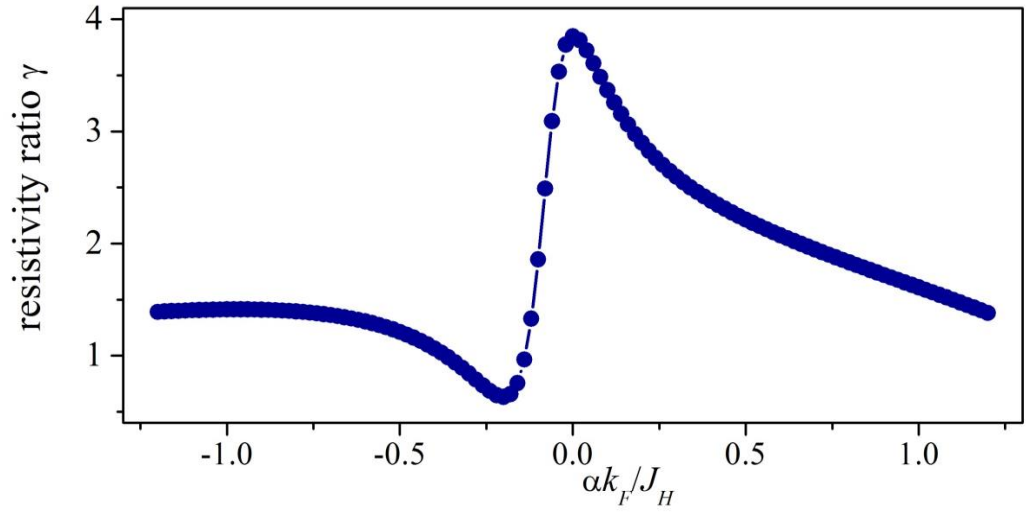

Fig. 4 (color online). Resistivity ratio as a function of $\alpha k_F/J_H$.

*To whom correspondence should be addressed: sxhuang@physics.miami.edu, jiadongzang@gmail.com, clc@pha.jhu.edu

#Present address: *Department of Physics, University of Miami, Coral Gables, FL 33146, USA*

References:

[1] M. N. Baibich *et al.*, Physical Review Letters **61**, 2472 (1988).
[2] G. Binasch *et al.*, Physical Review B **39**, 4828 (1989).
[3] C. H. Marrows, Advances in Physics **54**, 585 (2005).
[4] A. D. Kent *et al.*, Journal of Physics: Condensed Matter **13**, R461 (2001).
[5] P. M. Levy, and S. Zhang, Physical Review Letters **79**, 5110 (1997).
[6] J. F. Gregg *et al.*, Physical Review Letters **77**, 1580 (1996).
[7] R. P. van Gorkom, A. Brataas, and G. E. W. Bauer, Physical Review Letters **83**, 4401 (1999).
[8] G. Tatara, and H. Fukuyama, Physical Review Letters **78**, 3773 (1997).
[9] A. Aziz *et al.*, Physical Review Letters **97**, 206602 (2006).
[10] K. M. Seemann *et al.*, Physical Review Letters **108**, 077201 (2012).
[11] J. H. Franken *et al.*, Physical Review Letters **108**, 037205 (2012).
[12] Z. Yuan *et al.*, Physical Review Letters **109**, 267201 (2012).
[13] U. K. Roszler, A. N. Bogdanov, and C. Pfleiderer, Nature **442**, 797 (2006).
[14] S. Mühlbauer *et al.*, Science **323**, 915 (2009).
[15] X. Z. Yu *et al.*, Nature **465**, 901 (2010).
[16] I. Dzyaloshinsky, Journal of Physics and Chemistry of Solids **4**, 241 (1958).
[17] T. Moriya, Physical Review **120**, 91 (1960).
[18] P. Bak, and M. H. Jensen, Journal of Physics C: Solid State Physics **13**, L881 (1980).
[19] O. Nakanishi *et al.*, Solid State Communications **35**, 995 (1980).
[20] M. Uchida *et al.*, Science **311**, 359 (2006).
[21] N. Nagaosa, and Y. Tokura, Nat Nano **8**, 899 (2013).
[22] A. Fert, V. Cros, and J. Sampaio, Nat Nano **8**, 152 (2013).
[23] S. V. Grigoriev *et al.*, Physical Review B **76**, 224424 (2007).
[24] N. Manyala *et al.*, Nature **404**, 581 (2000).
[25] S. X. Huang *et al.*, submitted. (2014).
[26] Y. Onose *et al.*, Physical Review B **72**, 224431 (2005).